\documentclass[11pt]{article}
\usepackage{./fa2023}
\usepackage{amsmath}
\usepackage{cite}
\usepackage{url}
\usepackage{graphicx}
\usepackage{color}
\usepackage{siunitx}
\usepackage{balance}
\usepackage{arydshln}
\usepackage[utf8]{inputenc}

\title{Pre-trained Spatial Priors on Multichannel NMF for Music  Source Separation}

\multauthor
{Pablo Cabañas-Molero$^{1*}$ \hspace{1cm} Antonio J. Muñoz-Montoro$^2$ } { \bfseries{Julio Carabias-Orti$^1$ \hspace{1cm} Pedro Vera-Candeas$^1$}\\
$^1$ Departamento de Ingeniería de Telecomunicación, Universidad de Jaén, Linares, Spain\\
$^2$ Departamento de Ingeniería de Comunicaciones, Universidad de Malaga, Malaga, Spain
\correspondingauthor{pcabanas@ujaen.es}{P. Cabañas-Molero et al.}
}

\begin{document}

\maketitle
\begin{abstract}
This paper presents a novel approach to sound source separation that leverages spatial information obtained during the recording setup. Our method trains a spatial mixing filter using solo passages to capture information about the room impulse response and transducer response at each sensor location. This pre-trained
filter is then integrated into a multichannel non-negative matrix factorization (MNMF) scheme to better capture the variances of different sound sources. The recording setup used in our experiments is the typical setup for orchestra recordings, with a main microphone and a close ``cardioid'' or ``supercardioid'' microphone for each section of the orchestra. This makes the proposed method applicable to many existing recordings. Experiments on polyphonic ensembles demonstrate the effectiveness of the proposed framework in separating individual sound sources, improving performance compared to conventional MNMF methods.
\end{abstract}
%

\section{Introduction}\label{sec:introduction}

Classical music is typically recorded with all musicians performing together in the same room (unlike popular music which is often recorded one instrument at a time). Each instrument is captured by one or more close spots, usually cardioids aimed at the instrument, while the entire ensemble is captured by a stereo pair and other ambient microphones. The final recording is made by combining these various channels and adjusting their volume/delay to create a balanced stereo mix. However, since the close microphones also pick up sound from other instruments, the mixing work requires a certain effort to mask sound leakage. A source separation (SS) algorithm that can isolate individual instruments could significantly simplify this process. Also, having separated stems could be of interest for many applications such as 3D rendering, acoustic emphasis, acoustic scenarios recreation, and minus one.
 
In recent years, deep neural networks have made it possible to develop effective SS systems for popular music aided by the MUSDB dataset \cite{Rafii2018}. The objective of the demixing challenge is to address the issue of extracting vocals, bass, and drums from mixed and mastered pop songs. Regrettably, although most popular music can be recorded with individual takes for each performer using a reference metronome or backing track, ensembles typically record together in one take. This is because ensemble performers depend on hearing one another during the performance to achieve perfect synchronization. In fact, this lack of clean and sizeable real-world datasets for ensembles has limited the amount of research seen in this domain.

Several small ensembles databases have been presented for classical music source separation including URMP \cite{URMP2019}, Bach10 \cite{Duan2012Bach10}, or TRIOS \cite{Fritsch2012Trios} where the instruments are recorded individually and later realigned and downmixed to recreate a physical space for them. In this context, several deep learning approaches have been proposed including convolutional \cite{Miron2017} and recurrent \cite{Huang2014} layers which have the advantage of modeling a larger time context. Recently, several informed approaches have been presented including query-based method \cite{Seetharaman2019, Manilow2020a} for unseen (not pre-defined) sources, a semantic-based separation that incorporates auxiliary information such as score or video \cite{Gover2020, Slizovskaia2021}, and multitask settings \cite{Manilow2020}. Very recently, a larger dataset (6+ hours) of high-quality synthetic material has been presented in \cite{Sarkar2022}  with a range of string, wind, and brass instruments arranged as chamber ensembles that will enable novel approaches dealing with a higher number of instruments. In \cite{Tzinis} a single-channel separation system is trained by observing multichannel signals without the need of ground-truths, but is tested only on orthogonal sources (speech).

The aforementioned separation approaches are developed for single-channel mixtures. When multichannel signals are available, as is usual in orchestra recordings, separation can be improved by taking into account the spatial information of sources or the mixing process. In this context, multichannel Non-Negative Matrix Factorization (MNMF) \cite{Duong2010, Sawada2013, Ozerov2018, Sekiguchi2022} allows the separation of the individual sources (even in blind scenarios) while accounting for the spatial information such as inter-channel level difference (ILD) and inter-channel time difference (ITL). In fact, a recent comparison of blind MNMF w.r.t state-of-the-art single-channel deep learning solutions in \cite{Munoz2022} shows that exploiting spatial information allows competitive results even when training material is available for the networks. 


In the recording of orchestra music, before the performance, the audio engineer typically requests that each instrument plays solo for a few seconds. These segments are used later on by the engineer to align each spot with the stereo pair (delay compensation). In this paper, we investigate whether these preliminary solo recordings can be used to extract spatial priors for each source that can improve source separation. For our study, we use as a baseline the MNMF model proposed in \cite{Sawada2013} which is initialized using two types of priors: spatial information obtained from the preliminary recordings and instrument patterns from a pre-learned dictionary. The results we obtained from real and synthetic (created with a room simulator) mixtures indicate that integrating these priors into the MNMF model helps reduce sound leakage in close microphones in terms of objective BSS\_eval metrics.

\section{Background}

\subsection{Problem statement and assumptions}

We assume a recording with $M$ microphones and $S$ sources. Let $y_s(l)$ denote the time-domain signal played by source $s$. The observed mixture in microphone $m$ can be written as:
\begin{equation}
x_m(l) = \sum_{s=1}^S\sum_\tau h_{ms}(\tau) y_s(l-\tau),
\end{equation}
where $h_{ms}(\tau)$ is the impulse response from source $s$ to microphone $m$. We consider that impulse responses are time-invariant (static sources). In the frequency domain, the mixture can be expressed as:
\begin{equation}
\mathbf{x}_{ft} = \sum_{s=1}^S \mathbf{h}_{fs} y_{sft},
\end{equation}
where $\mathbf{x}_{ft} = [x_{1ft}, \dots, x_{Mft}]^T$ is the observed $M$-channel complex spectrogram in frequency $f$ and frame $t$, and $\mathbf{h}_{fs} = [h_{1fs}, \dots, h_{Mfs}]^T$ is the mixing filter from source $s$ to the $M$ microphones.

We assume that each source is recorded with at least one close microphone (spot), which necessarily implies that $M \geq S$. Without any loss of generality, let us denote that channel $m = s$ is a close spot of source $s$. The problem addressed in this work is to retrieve the image of each source in its close spot 
given the $M$-channel mixture $\mathbf{x}_{ft}$.

\subsection{Multichannel NMF}

In MNMF the non-negative 
spectrogram of the sources is modeled with NMF, while the complex spectrogram of the $M$-channel mixture is modeled using complex Gaussian distributions. Let $\mathbf{X}_{ft}$ be the $M \times M$ spatial covariance matrix (SCM) computed from the observed mixture:
\begin{equation}
\mathbf{X}_{ft} = \begin{bmatrix}
	\left|x_{1ft}\right| & \dots & x_{1ft}x_{Mft}^* \\
	\vdots & \ddots & \vdots \\
	x_{Mft}x_{1ft}^* & \dots & \left|x_{Mft}\right|
\end{bmatrix},
\end{equation}
%
%
%
where
$(\cdot)^*$ is complex conjugate. It can be observed that this matrix encodes spatial information in its off-diagonal, in the form of magnitude cross correlation and phase difference.

Similarly to \cite{Sawada2013}, a $M \times M$ complex matrix $\mathbf{H}_{fs}$ can be defined to model the spatial property of source $s$. The observed SCM can be modeled following a NMF scheme as:
\begin{equation}\label{eq:model}
\hat{\mathbf{X}}_{ft} = \sum_{s=1}^S \mathbf{H}_{fs} \sum_{k=1}^K t_{sfk} v_{skt},
\end{equation}
where $t_{sfk}$ is the $k$-th basis of source $s$ in frequency $f$, and $v_{skt}$ is the gain of this basis at frame $t$, both real and non-negative. Note that each source is modeled using $K$ bases.

By formulating appropriate statistical models, \cite{Sawada2013} derives multiplicative update rules to minimize the divergence between the observed SCM $\mathbf{X}_{ft}$ and its modeled form $\hat{\mathbf{X}}_{ft}$. In the case of Euclidean divergence, the rules to update $\mathbf{H}_{fs}$ and $v_{skt}$ are:
\begin{equation}\label{eq:updateV}
v_{skt} \leftarrow v_{skt} \frac{\sum_{f} t_{sfk} \mathtt{tr}(\mathbf{X}_{ft} \mathbf{H}_{fs})}
{\sum_{f} t_{sfk} \mathtt{tr}(\hat{\mathbf{X}}_{ft} \mathbf{H}_{fs})},
\end{equation}
\begin{equation}\label{eq:updateH}
	\begin{aligned}
\mathbf{H}_{fs} \leftarrow \mathbf{H}_{fs} & \left( \sum_{ftk} t_{sfk} v_{skt} \hat{\mathbf{X}}_{ft} \right)^{-1} \\
 \times & \left( \sum_{ftk} t_{sfk} v_{skt} \mathbf{X}_{ft} \right),
	\end{aligned}
\end{equation}
where $\mathtt{tr}(\cdot)$ is the trace of a square matrix. After each iteration, some post-processing is required to keep $\mathbf{H}_{fs}$ Hermitian and positive semidefinite \cite{Sawada2013}.

\section{Proposed system based on priors}

Our approach involves two stages. In the first stage, we estimate the mixing filters $\mathbf{H}_{fs}$ by using the preliminary solo passages recorded by the audio engineer before the concert. These short passages are manually annotated in a text file using time markers. Since each source is free of interference here, our system can capture accurate spatial information for each instrument. The only requirement is that each solo segment contains signal energy in each frequency point $f$ relevant for the instrument. In the second stage, focused on source separation during the performance, we estimate the time-varying gains $v_{skt}$ using two fixed priors: the spatial filters $\mathbf{H}_{fs}$ and a dictionary of instrument bases $t_{sfk}$.

\subsection{Spatial priors estimation}

Let $\mathbf{x}_{ft}^{(s)}$ be the $M$-channel observation for a time-frequency point of the solo segment of source $s$. Assuming that its spot microphone $m=s$ is very close to the instrument, we decide to ignore the path between source and spot (i.e., $h_{mfs}=1$, when $m=s$) and approximate the source magnitude by its spot image: $|y_{sft}| \approx |x_{mft}^{(s)}|$, for $m=s$. In this case, we can write a single-source version of the model in \eqnref{eq:model} as:
\begin{equation}\label{eq:modelsingle}
	\hat{\mathbf{X}}_{ft} = \mathbf{H}_{fs} \left|x_{sft}^{(s)} \right|.
\end{equation}
This model approximates the observed SCM from the spot spectrogram magnitude, under the assumption that instrument and spot are in the same spatial location.

The unknown of the model, the spatial filter $\mathbf{H}_{fs}$, is initialized to real random values with 1 mean, and updated iteratively as in \eqnref{eq:updateH}, which in this case is reduced to:
\begin{equation}
		\mathbf{H}_{fs} \leftarrow \mathbf{H}_{fs} \left( \sum_{ft} \left| x_{sft}^{(s)} \right| \hat{\mathbf{X}}_{ft} \right)^{-1}
		\left( \sum_{ft} \left| x_{sft}^{(s)} \right| \mathbf{X}_{ft} \right),
\end{equation}
We observed that the algorithm converges very fast to a local minima, in less than 5 iterations. The process is repeated for each instrument $s$, resulting in a set of $S$ spatial filters.

\subsection{Instrument priors}

For each classical instrument, we pre-trained a dictionary of bases using the RWC instrumental database. A single basis was learned for each playable note/pitch. To do this, we ran a single-channel 1-basis NMF model on the magnitude spectrogram of each individual note, using $\beta$-divergence with $\beta=1.5$ and 50 iterations. $K = 115$ bases are reserved for each instrument, with zero values used to pad the instrument if it has fewer notes.

\subsection{Source separation}

In the separation stage, the signal model in \eqnref{eq:model} is set with the spatial priors $\mathbf{H}_{fs}$ learned in the solo sections, and with the bases $t_{sfk}$ of the instruments involved in the mixture. The free parameters of the model are the time-varying gains $v_{skt}$. They are initialized with 1 mean random values, and updated using \eqnref{eq:updateV}. We ran the algorithm with 50 iterations.

Finally, each source is reconstructed using the estimated model parameters and its close spot. To ensure a conservative reconstruction, we use Wiener masks, which represent the energy proportion of each source w.r.t the total energy of the mixture. The estimated source $\hat{y}_{sft}$ is retrieved by applying the single-channel Wiener mask on spot $m=s$, as follows:
\begin{equation}
\hat{y}_{sft} = \frac{ [\mathbf{H}_{fs}]_{mm} \sum_{k=1}^K t_{sfk} v_{skt} }
{\sum_{s=1}^S [\mathbf{H}_{fs}]_{mm} \sum_{k=1}^K t_{sfk} v_{skt}} x_{mft}, \quad m=s.
\end{equation}
Here, $[\cdot]_{mm}$ means matrix indexing. Elements $[\mathbf{H}_{fs}]_{mm}$ are always real, since they approximate the diagonal of the observed SCM. Also, it can be observed that the estimated $\hat{y}_{sft}$ has the same phase than $x_{mft}$.

\section{Experiments and results}

\subsection{Datasets}

We tested our system on ensemble recordings with $S=4$ instrument/sections. We employed 2 datasets: real and simulated. The \emph{real dataset} was recorded in a professional studio using a multi-microphone setup, including a cardioid spot for each section and a main pair. Four sections were recorded: Violin 1, Viola, Cello and Bass. Each section was separately recorded (musicians used a 3rd party performance as a reference), an then aligned on the main pair to assemble the mixture. Consequently, we have access to the ground-truth images of each section in all microphones. The dataset includes passages of Mozart's KV550 and Tchaikovsky's Romeo and Juliet, resulting in almost 3 minutes of audio divided into 4 tracks.

The \emph{simulated dataset} was created using 4 pieces from the URMP database \cite{li2018}. Multichannel mixtures were simulated (using the image method) in a room with size $22 \times 16 \times 5$ \mbox{m$^3$}, reverberation \mbox{RT$_{60}=1$ s}, and with omnidirectional spots, assuming musicians separated 2 m each other. This dataset includes wind, string and mixed ensembles, and comprises around 4 minutes of audio.

A preliminary solo passage was generated for each source. In the real dataset, each section plays a full scale for 40 seconds. In the simulated dataset, the solo consists in 3 hand clapping sounds in the instrument position.


\subsection{Experimental setup}
We used a STFT computed with frame-length 2048 samples (at 44,1 kHz), half frame overlap and half-sine windows. We tested three variants of the algorithm: 
\begin{itemize}\setlength\itemsep{-0.25em}
    \item \emph{MNMF fixH}, with fixed priors.
    \item \emph{MNMF initH}, that updates prior $H_{fs}$.
    \item \emph{MNMF}, without $H_{fs}$ prior (random initialization).
\end{itemize}
For reference, we compared our results with 2 methods: 
\begin{itemize}\setlength\itemsep{-0.25em}
    \item \emph{NTF Panning} (non-negative tensor factorization) in \cite{cabanasJOS}, using panning coefficients learned in the solo segments.
    \item \emph{FastMNMF2} \cite{Sekiguchi2022}, that was executed blind (no priors) with the code released by the authors.
\end{itemize}
All methods worked only on the $M=4$ spot channels. Results were obtained using the BSS\_eval toolbox. 

\subsection{Results}
In real mixtures (\tabref{tab:real}), the best results are obtained by \emph{MNMF fixH} ($\Delta \text{SDR}=4.9$ dB), closely followed by \emph{NTF Panning}. This suggests that \emph{MNMF fixH} is mainly exploiting inter-channel amplitude differences in this scenario. Interestingly, we observed that updating the spatial priors causes a severe performance drop. \emph{MNMF} is unable to separate the sources without priors in this case. \mbox{\emph{FastMNMF2}}, on the contrary, is indeed able to separate sources, but still performs worse than \emph{MNMF fixH}.

\begin{table}[!h]
 \caption{Results on real mixtures (dB).}
 \begin{center}
 \begin{tabular}{lrrr}
  \hline
  \hline
  \bf{Method} & \bf{$\Delta$SDR} & \bf{$\Delta$SIR} & \bf{SAR} \\
  \hline  
  MNMF fixH   & 4.90  & 7.19 & 13.31 \\
  MNMF updtH  & 1.92  & 7.10 & 8.16 \\
  MNMF        & -0.26 & 4.98 & 5.91 \\
  \hdashline
  FastMNMF2   & 2.29  & 5.12 & 10.50 \\
  \hdashline
  NTF Panning & 4.36  & 6.83 & 12.97 \\
  \hline
  \hline
 \end{tabular}
\end{center}
 \label{tab:real}
\end{table}

In simulated mixtures (\tabref{tab:sim}), \emph{MNMF fixH} performs significantly better than \emph{NTF Panning} ($7.32$ vs $5.68$ dB). This can be due to the fact that, in a simulated scenario with point sources, the MNMF model is able to fit the spatial mixing more accurately, exploiting both amplitude and phase differences. \emph{MNMF fixH} outperformed the other systems in this scenario as well.

\begin{table}[!h]
 \caption{Results on simulated mixtures (dB).}
 \begin{center}
 \begin{tabular}{lrrr}
  \hline
  \hline
  \bf{Method} & \bf{$\Delta$SDR} & \bf{$\Delta$SIR} & \bf{SAR} \\
  \hline
  MNMF fixH   & 7.32 & 8.80 & 10.26 \\
  MNMF updtH  & 4.02 & 6.04 & 6.34 \\
  MNMF        & 2.84 & 4.72 & 4.85 \\
  \hdashline
  FastMNMF2   & 3.78 & 7.12 & 6.69 \\
  \hdashline
  NTF Panning & 5.68 & 7.61 & 8.91 \\
  \hline
  \hline
 \end{tabular}
\end{center}
 \label{tab:sim}
\end{table}

\section{Conclusions}

This paper presents a novel approach to music source separation in orchestra recordings by leveraging spatial information obtained during the recording setup. The method combines a pre-trained spatial mixing filter with a MNMF scheme. The results demonstrate the effectiveness of the proposed approach, achieving the best performance in both real and simulated mixtures. It successfully utilizes inter-channel amplitude differences and accurately fits the spatial mixing in simulated scenarios with point sources.

The practical implications of this research are important for classical music recording and source separation. By effectively separating individual sound sources, the proposed method simplifies the mixing process and reduces the need for extensive sound leakage masking. Additionally, the availability of separated stems enables various applications such as 3D rendering, acoustic emphasis, acoustic scenario recreation, and creating minus one tracks.

It is important to acknowledge that while the proposed method demonstrates promising results, there is still room for further improvement. Future research can explore additional techniques to better incorporate and exploit spatial information, taking into account factors such as room impulse responses and transducer responses. Moreover, the availability of larger and more diverse datasets specifically tailored to ensemble recordings would facilitate the development of more advanced and robust source separation algorithms.

\section{Acknowledgments}
This work was supported by the Regional Government of Andalucía under the PAIDI 2020 Framework Programme through grant P18-TPJ-4864 and the REPERTORIUM project. Grant agreement number 101095065. Horizon Europe. Cluster II. Culture, Creativity and Inclusive society. Call HORIZON-CL2-2022-HERITAGE-01-02.

\bibliography{fa2023_template}

\begin{thebibliography}{10}

\bibitem{Rafii2018}
Z.~Rafii, A.~Liutkus, F.-R. St{\"o}ter, S.~I. Mimilakis, and R.~Bittner, ``The
  {MUSDB18} corpus for music separation,'' Dec. 2017.

\bibitem{URMP2019}
B.~Li, X.~Liu, K.~Dinesh, Z.~Duan, and G.~Sharma, ``Creating a multitrack
  classical music performance dataset for multimodal music analysis:
  Challenges, insights, and applications,'' {\em IEEE Transactions on
  Multimedia}, vol.~21, no.~2, pp.~522--535, 2019.

\bibitem{Duan2012Bach10}
Z.~Duan and B.~Pardo, ``Bach 10 dataset---- a versatile polyphonic music
  dataset,'' 2012.

\bibitem{Fritsch2012Trios}
J.~Fritsch, ``High quality musical audio source separation. master’s
  thesis,'' Master's thesis, UPMC / IRCAM / Telecom Paristech, 2012.

\bibitem{Miron2017}
M.~Miron, J.~Janer, and E.~G{\'o}mez, ``Monaural score-informed source
  separation for classical music using convolutional neural networks,'' in {\em
  International Society for Music Information Retrieval Conference}, 2017.

\bibitem{Huang2014}
P.-S. Huang, M.~Kim, M.~Hasegawa-Johnson, and P.~Smaragdis, ``Deep learning for
  monaural speech separation,'' in {\em 2014 IEEE International Conference on
  Acoustics, Speech and Signal Processing (ICASSP)}, pp.~1562--1566, 2014.

\bibitem{Seetharaman2019}
P.~Seetharaman, G.~Wichern, S.~Venkataramani, and J.~Le~Roux,
  ``Class-conditional embeddings for music source separation,'' pp.~301--305,
  05 2019.

\bibitem{Manilow2020a}
E.~Manilow, G.~Wichern, and J.~L. Roux, ``Hierarchical musical instrument
  separation,'' in {\em International Society for Music Information Retrieval
  Conference}, 2020.

\bibitem{Gover2020}
M.~Gover and P.~Depalle, ``Score-informed source separation of choral music,''
  in {\em International Society for Music Information Retrieval Conference},
  2020.

\bibitem{Slizovskaia2021}
O.~Slizovskaia, G.~Haro, and E.~Gómez, ``Conditioned source separation for
  musical instrument performances,'' {\em IEEE/ACM Transactions on Audio,
  Speech, and Language Processing}, vol.~29, pp.~2083--2095, 2021.

\bibitem{Manilow2020}
E.~Manilow, P.~Seetharaman, and B.~Pardo, ``Simultaneous separation and
  transcription of mixtures with multiple polyphonic and percussive
  instruments,'' pp.~771--775, 05 2020.

\bibitem{Sarkar2022}
S.~Sarkar, E.~Benetos, and M.~Sandler, ``Ensembleset: A new high-quality
  synthesised dataset for chamber ensemble separation,'' in {\em International
  Society for Music Information Retrieval Conference}, 2022.

\bibitem{Tzinis}
E.~Tzinis, S.~Venkataramani, and P.~Smaragdis, ``Unsupervised deep clustering
  for source separation: Direct learning from mixtures using spatial
  information,'' in {\em ICASSP 2019}, pp.~81--85, 2019.

\bibitem{Duong2010}
N.~Q.~K. Duong, E.~Vincent, and R.~Gribonval, ``Under-determined reverberant
  audio source separation using a full-rank spatial covariance model,'' {\em
  IEEE Transactions on Audio, Speech, and Language Processing}, vol.~18, no.~7,
  pp.~1830--1840, 2010.

\bibitem{Sawada2013}
H.~Sawada, H.~Kameoka, S.~Araki, and N.~Ueda, ``Multichannel extensions of
  non-negative matrix factorization with complex-valued data,'' {\em IEEE
  Transactions on Audio, Speech, and Language Processing}, vol.~21, no.~5,
  pp.~971--982, 2013.

\bibitem{Ozerov2018}
A.~Ozerov, C.~F{\'e}votte, and E.~Vincent, {\em An Introduction to Multichannel
  NMF for Audio Source Separation}.
\newblock Cham: Springer International Publishing, 2018.

\bibitem{Sekiguchi2022}
K.~Sekiguchi, Y.~Bando, A.~A. Nugraha, K.~Yoshii, and T.~Kawahara, ``Fast
  multichannel nonnegative matrix factorization with directivity-aware
  jointly-diagonalizable spatial covariance matrices for blind source
  separation,'' {\em IEEE/ACM Transactions on Audio, Speech, and Language
  Processing}, vol.~28, pp.~2610--2625, 2020.

\bibitem{Munoz2022}
A.~J. Mu{\~n}oz-Montoro, J.~J. Carabias-Orti, P.~Caba{\~n}as-Molero, F.~J.
  Ca{\~n}adas-Quesada, and N.~Ruiz-Reyes, ``Multichannel blind music source
  separation using directivity-aware mnmf with harmonicity constraints,'' {\em
  IEEE Access}, vol.~10, pp.~17781--17795, 2022.

\bibitem{li2018}
B.~Li, X.~Liu, K.~Dinesh, Z.~Duan, and G.~Sharma, ``Creating a multitrack
  classical music performance dataset for multimodal music analysis:
  Challenges, insights, and applications,'' {\em IEEE Transactions on
  Multimedia}, vol.~21, no.~2, pp.~522--535, 2018.

\bibitem{cabanasJOS}
P.~Caba{\~{n}}as-Molero, A.~J. Mu{\~{n}}oz-Montoro, P.~Vera-Candeas, and
  J.~Ranilla, ``The music demixing machine: toward real-time remixing of
  classical music,'' {\em The Journal of Supercomputing}, Apr 2023.

\end{thebibliography}
\balance
%
%
%

\end{document}